\documentclass[angeod,manuscript]{copernicus}  %82 pp a 1500
\nolinenumbers
\usepackage{amsmath}
\usepackage{color}
\usepackage{babel}
\usepackage[T1]{fontenc}

\usepackage{natbib}
\newcommand{\MC}{\mathcal}

\begin{document}

\title{A Note on the Entropy Force in Kinetic Theory and Black Holes
}

\author[1,3]{Rudolf A. Treumann}
\author[2]{Wolfgang Baumjohann}
%\author[2]{Yasuhito Narita}
\affil[1]{International Space Science Institute, Bern, Switzerland}
\affil[2]{Space Research Institute, Austrian Academy of Sciences, Graz, Austria}
\affil[3]{Geophysics Department, Ludwig-Maximilians-University Munich, Germany\\

\emph{Correspondence to}: Wolfgang.Baumjohann@oeaw.ac.at
}

\runningtitle{Reconnection theory}

\runningauthor{R. A. Treumann \& W. Baumjohann}

\received{ }
\pubdiscuss{ } %% only important for two-stage journals
\revised{ }
\accepted{ }
\published{ }

%% These dates will be inserted by the Publication Production Office during the typesetting process.

\firstpage{1}

\maketitle

\begin{abstract}
%\noindent\textbf{Abstract}. -- 
The entropy force is the collective effect of inhomogeneity in disorder in a statistical many particle system. We demonstrate its presumable effect on one particular astrophysical object, the black hole. We then derive the kinetic equations of a large system of particles including the entropy force. It adds a collective therefore integral term to the Klimontovich equation for the evolution of the one-particle distribution function. Its integral character transforms the basic one particle kinetic equation into an integro-differential equation already on the elementary level, showing that not only the microscopic forces but the hole system reacts to its evolution of its probability distribution in a holistic way. It also causes a collisionless dissipative term which however is small in the inverse particle number and thus negligible. However it contributes an entropic collisional dissipation term. The latter is defined via the particle correlations but lacks any singularities and thus is large scale. It allows also for the derivation of a kinetic equation for the entropy density in phase space. This turns out to be of same structure as the equation for the phase space density. The entropy density determines itself holistically via the integral entropy force thus providing a self-controlled evolution of entropy in phase space. 
\keywords{Collisionless reconnection, electron scale current sheets, solar wind}

%\vspace{0.5cm}
%\noindent\textbf{Abstract}.-- 
%\abstract 
%\vspace{1cm}
\end{abstract}
\section{Introduction: Entropy Force}
About thirty years ago, \citet{prigogine1980} attempted a microscopic theory of entropy assuming that, by some quantum process, seeds of entropy could be generated. Such a hypothetical process would, in the early universe, possibly lay down the direction of time. Unfortunately, so far, such microscopic sources of entropy have not been confirmed. It seems that they can hardly be expected because quantum uncertainty itself is a stochastic process, which by its own nature does not contain any direction. It is hard to believe that it could lead to entropy production if not aided by some kind of dissipative interaction. Entropy is a thermodynamic concept, which by itself requires an underlying dynamics, which allows for the presence of many states that a system consisting of many subsystems, components, particles would be able to occupy. 

More recently, it has been speculated \citep{verlinde2011} that that kind of a mesoscopic entropy in quantum string theory could cause gravity to emerge from the action of a quantum entropic force as a gradient of entropy generated in string interactions, intended to provide a physical basis for the so-called modified Newtonian gravity, which proposes that Newton's law should be corrected on the large scales to eliminate the problem of dark matter in astronomy. 

From a completely different point of view, the idea of an entropy force has been picked up in the discussion of maximum entropy methods in prediction theory \citep{wissner2013,kappen2013} in open systems where the probabilistic version of entropy depends on space and time, propagates into the future and, thus, has a finite gradient in space and time, which is interpreted as force. It apparently is capable of allowing, based~on maximization of entropy, predicting the time evolution of the system, an~interesting and possibly far-reaching predictive concept. In a sufficiently small closed system, it necessarily must describe the evolution of entropy towards a finite thermal state of maximum entropy. Recently,~entropy~forces have also been applied to molecular dynamics in proteins (\citep{keul2018} and the references therein). Spatial smallness is required by causality to enable synchronization. Therefore, the concept applies to the universe on the cosmological timescale only in order to allow for homogenization of entropy. On~any local scale, the entropy produced in the classical system represents a localized excess in entropy. If not artificially confined, this excess tends to expand and affect its environment. This necessarily generates a local entropy force, a classical force that should not be mixed up with the above-mentioned entropic force in string systems. This force follows from the first law in thermodynamics: 
\begin{equation}
dE=TdS-PdV
\end{equation}
which relates the three different forms of energy $E$, pressure $PV$, and entropy $TS$. Gradients in energy, pressure, and temperature are known to be forces, and the gradient of volume causes dispersion, flows, and forces, for instance in charged systems. 

In a similar vein, a gradient in entropy corresponds to a collective macroscopic effect as the entropy tends to expand and maximize. This is a purely macroscopic effect indeed because, similar to density/volume and pressure, the entropy $S$ is defined only for macroscopic systems, consisting of a large number of subsystems, to which finite temperature and density can be assigned and which occupy a finite volume. In the first law, it is only the energy $E$ that maintains its meaning also in the microscopic world down to only one particle, to which assigning temperature makes no sense. The entropy potential $U=TS$ indeed is not just a thermodynamic potential; it is also a real potential always being positive and thus repulsive. The entropy force is then given as its gradient: 
\begin{equation}
\mathbf{F}=-\nabla U
\end{equation}
as usually taken negative. It consist of two parts, a thermal force $ -S\nabla T$, which is of no interest here, and the genuine entropy force: 
\begin{equation}
\mathbf{F}_S=-T\nabla S
\end{equation}
This might look trivial; however, it is not, as we will demonstrate below with a particular example: the black hole.

However, before proceeding, we recall that, since both $T$ and $S$ are positive definite, the entropy force is repulsive in the direction negative to the gradient of entropy. This means that an accumulation of entropy at some location, if not artificially confined to a box, will act outward. Adopting an interpretation of entropy as disorder, which by no means is generally justified, thus implies that disorder tends to infect its external region. It has the tendency to expand. 

With temperature $T$ in energy units, the entropy $S$ has no dimension. Moreover, the product of temperature and entropy is a scalar function with the dimension of a potential. For scalar temperature, i.e., at temperature isotropy, $S$ is also a scalar. Under conditions of anisotropic temperature, the inverse temperature becomes a vector \citep{nakamura2009}, and thus, $S$ becomes a vector as well (more generally, both become tensors). In the interest of simplicity, we do not consider this case in the following.

This entropy force does {not depend} on particle mass or charge, at least not explicitly. Mass is contained in temperature and energy, but there is no explicit reference to it in the definition of the entropy force. Thus, {for a given temperature, all particles independent of their properties will be subject to the same entropy force}. In this sense, the entropy force is a general mechanical force seeking to {restore smoothness in disorder} on a higher level of disorder, completely independent of which kind of particles have contributed to the inhomogeneity in disorder.

By its nature, the above entropy force is a long-range force. It does not compete with the Coulomb force on the short scales. From the first law, one realizes that the main force it competes with is the pressure gradient. Both are proportional to the temperature, which thus drops out when comparing the forces. Both are proportional to the density gradient. Thus, in the presence of changing volume, the entropy force adds to or diminishes the effect of the pressure force. 

%%%%%%%%%%%%%%%%%%%%%%%%%%%%%%%%%%%%%%%%%%
\section{Entropy Force of Schwarzschild Black Holes}

Let us turn to our example, the Schwarzschild black hole, which we chose for demonstration because of its simplicity and cleanness compared to Kerr or charged Nordstrom black holes. Black~holes are known to carry entropy \citep{bekenstein1972,bekenstein1973,bardeen1973,hawking1971,hawking1975,hawking1976}. More correctly, since the interior of the black hole is not accessible, it is the black hole horizon that carries an entropy. This came as a surprise, as it implies that the horizon possesses a temperature and therefore must be considered as a macrosystem, which occupies a large number of states. Microscopically, this puzzle has not been resolved until today, even though a large number of attempts have been put forward to elucidate the internal structure of the horizon (see\citep{brout1995,jacobson1995,doran2008,padmanabhan2009,taylor2013} and several others). 
Such considerations were based on the Bekenstein--Hawking entropy and the Hawking radiation of a black hole \citep{hawking1975}, which is attributed to its finite entropy and thus finite temperature. It implies the existence of a thermodynamic for the black hole horizon with the implication that the horizon physics involves a very large number of states that can be occupied. \citet{jacobson1995} extended this concept to horizons in general in order to develop a thermodynamics of gravitational horizons from which he found that Einstein's gravitational field equations formally play the role of equations of state. This concept was reviewed and extended subsequently \mbox{by~\citet{padmanabhan2009}} to speculate about the general importance of horizon physics in general relativity and cosmology, suggesting that all the physics is holographically contained in the physics of horizons. 

Schwarzschild black holes are in the first place classical objects. However, their entropy includes the quantum nature of matter at the horizon (cf., e.g., \citep{brout1995}), which is induced by the sharpness of the horizon and indicates that black holes are not purely classical. Let us ask what the entropy force related to the presence of the horizon would be.

%%%%%%%%%%%%%%%%%%%%%%%%%%%%%%%%%%%%%%%%%%
\subsection{Schwarzschild Constant}

A Schwarzschild black hole of mass $M$ has energy $Mc^2$, radius $R_S =2GM/c^2$, and spherical surface $A_{BH}=4\pi R_S^2$. Forming the ratio of the total black hole energy and the Schwarzschild radius~yields:
\begin{equation}
\MC{F}_S=Mc^2/R_S=c^4/2G=6.053\times 10^{\:43}\quad\mathrm{N}
\end{equation} 
a constant that has the dimension of a force and that we call the {Schwarzschild constant}. This~is a universal constant, whose value is independent of any property of the black hole, a force. This~force is the Planck force, which so far has not been given any physical meaning. We prefer to call it the Schwarzschild constant as the black hole is the only place where it naturally arises.

\subsection{Horizon Entropy Force}
The entropy of the black hole horizon is the (dimensionless) Bekenstein--Hawking entropy: 
\begin{equation}
S_{BH}=\frac{A_{BH}}{4\lambda_P^2}=\frac{\pi R_S^2c^3}{G\hbar}
\end{equation}
with $\lambda_P$ the Planck length, and its corresponding black-body radiation temperature is the Hawking~temperature: 
\begin{equation}
T^{BH}=k_BT_H=\hbar c^3/8\pi GM
\end{equation}
here given in energy units. This yields trivially the Bekenstein--Hawking energy ${E}_{BH}=T^{BH}S_{BH}=\frac{1}{2}Mc^2$ of the horizon, just half the black hole energy. For a classical black hole, the horizon has no width, suggesting an infinite gradient when crossing it. In order to obtain the force, the relevant distance taken for the gradient is the diameter $2R_S$ of the black hole, yielding for the modulus of the outward directed entropy force:
\begin{equation}
F_S^{BH}\sim\ \frac{T^{BH}S_{BH}}{2R_S} = \frac{c^4}{8G} = \frac{\MC{F}_S}{4} \approx 1.5\times 10^{\,43} ~\mathrm{N}
\end{equation}
a quarter of the {Schwarzschild constant}. A more precise calculation would correct for the numerical factor, which, however, is of order $O(1)$. Formally, the Schwarzschild constant, and thus also the entropy force at the horizon, is in fact a very strong force. Its presence, if real, at the horizon is rather surprising. It suggests that the physics of what happens inside the black hole is not really well enough~understood. 

Forming the ratio of the gravitational force $F_G^{BH}= -GmM/R_S^2$ any particle of mass $m$ experiences when approaching and touching the black hole horizon to this horizon entropy force, one obtains:
\begin{equation}
\Big|F_G^{BH}/F_S^{BH}\Big|\sim m/M
\end{equation}
For any massive black hole and any normal mass particle, this is a small number. A light particle $m\ll M$ will barely overcome this repulsion when hitting the horizon. To overcome it, it requires the collision of two black holes of nearly equal mass $M_1\sim M_2$, which would make the ratio $m/M\to M_1/M_2\approx 1$. Collisions of such nearly-equal mass black holes have only recently been detected by the spectacular observation of gravitational waves.

{At its horizon,} the entropy force of a massive black hole $M\gg m$ compensates by far for the black hole's gravitational attraction on $m$. This is a consequence of the enormous sharpness of the entropy gradient at the horizon where the entropy is restricted to the surface of the horizon only, whose width is not precisely given, but as generally assumed, is of the order of a few Planck lengths $\lambda_P$ only. This force is remarkable only at the horizon itself when the mass $m$ gets into contact with the horizon. It will not be susceptible at some larger finite distance. 

This follows from the fact that there is no known classical entropy field that would allow the entropy force to extend a distance ahead of the black hole into the surrounding space and is conjectured from the complete absence of any radial dependence of the force outside the horizon. In~this picture, the entropy gradient is felt only locally across the horizon of the black hole when the particle touches it, an instant that is never seen or experienced by an external observer for whom the time the particle approaches the horizon stretches out to infinity. The particle, however, does in fact experience the presence of the black hole and, assuming that it remains intact having survived the enormous attraction during its inward spiraling motion, in its proper frame at proper time, really~touches the horizon and wants to cross it. Shortly before this instant, however, the horizon entropy comes into play and stops the particle.

The gravitational force on the particle of mass $m$ (assuming it retains its mass till reaching the horizon, which is certainly not the case) would overcome the entropic force still only at a small fraction of the radius given by: 
\begin{equation}
\frac{\Delta r}{R_S}\approx \bigg(\frac{M_\odot}{M}\bigg)^\frac{1}{2}\bigg(\frac{m}{M_\odot}\bigg)^\frac{1}{2}\approx \ 3\times10^{-28} \bigg(\frac{M_\odot}{M}\bigg)^\frac{1}{2}
\end{equation}
where on the right, we assumed a proton. For instance, this distance for a proton and a $M=10^8M_\odot$ massive black hole is of the order of only $\Delta r\sim 10^{-23}$ m, deep inside the submicroscopic domain, though ten orders of magnitude larger than the Planck length. If it applies, then it would cause accumulation of matter in a film of roughly this width only. 

The classical picture does not inform about the microscopic physics going on when this happens. Elucidating the real physics requires a quantum electrodynamic calculation for instance along the paths drawn by Hawking when calculating the black-body black hole radiation. Referring to Hawking's results implies that the horizon will be surrounded by a dilute and thin radial dust film of newly- and continuously-created virtual particles, which sustain and support the weak Hawking radiation when tunneling into reality. The radial extension of this dust film implies a softening of the radial entropy gradient corresponding to a finite radial extension of the action region of the entropy force. It would be this radial domain where the light particle in its inward spiraling motion becomes trapped and retarded and is ultimately stopped and prevented from entering the interior of the black hole. 

\subsection{Body Entropy}

What happens to the entropy inside the black hole would be important to know (cf. \citep{hamilton2018} for a discussion of Hawking radiation inside black hole geometry) in order to resolve the puzzle. It is rather improbable that the entropy would be constant inside the black hole, as this would require that the interior is in thermal equilibrium, which it is certainly not when being under the conditions of collapsing matter under gravitational attraction. One might speculate that towards deeper inside, the entropy decreases with decreasing radius, because the surface decreases as $\sim$$(r/R_S)^2$. Assigning a Hawking temperature to each shell of such a radius, the corresponding Hawking temperature would increase only as $\sim$$R_S/r$. Thus any entropy-force potential should decrease towards the interior like $\sim$$r/R_S$. The outer black hole horizon becomes the black hole shell of maximum entropy, the~radius where the black hole entropy maximizes. The interior entropy force, the gradient of the potential, remains constant throughout the entire interior volume of the black hole with the exclusion of the~singularity. 

Notably, this entropy force points towards the interior of the hole, i.e., towards the singularity. It~thus adds to the already existing gravitational acceleration being felt throughout the entire interior. By pointing inside towards decreasing radius $r$, it would push any existing massive particle that made it across the horizon into the singularity up in energy. This probably means that classically, no massive particle can make it across the horizon. It can only be non-massive radiation that crosses inward: photons and gluons, the massless bosons of electrodynamics and chromodynamics. Whether massive particles like electrons and quarks can indeed tunnel across the horizon remains a question that cannot be answered in the realm of classical physics.

Admittedly, these considerations are rather speculative as long as the evolution of entropy with increasing radial distance from the horizon towards inside and also outside the black hole has not been microscopically inferred. In any case, the question of the horizon representing a sharp surface remains a question that probably only quantum gravity can give an ultimate answer to, as it must proceed on scales close to the Planck scale $\lambda_P$. This is not our concern here.

\subsection{Visibility and Matter Digestion}
Naked isolated black holes are invisible except for their weak and so far inaccessible Hawking radiation. The question why black holes, which are embedded into surrounding matter, when~accreting become visible at all is comparably easy to answer. Any mass flow approaching the horizon before encounter feels the gravitational field of the black hole, spirals in, accelerates, heats up, becomes~partially transformed into radiation, and starts radiating violently. General relativity indicates that this process stretches time to infinity. Hence, even though the matter starts as material particles that cannot be digested by the black hole, because matter cannot pass the horizon due to the barrier the classical entropic force provides for massive particles, the emitted radiation is visible for long. The~latter is a well-known fact, and though radiation from accreting black hole suspects has been observed for decades already, the observational proof has only very recently been given when a particular black hole signature could be resolved in radio emission.

During inward spiraling, the matter irradiates, which happens for all the matter that consists of much smaller mass particles (gaseous clouds, dust, stars) than the black hole. The total mass of the matter that hits the horizon at each instant is much less than $M$. The radiation produced in this process may be considered isotropic because there is no remarkable beaming until the matter becomes charged during gravitational compression, heating, and ionization. It then via a dynamo process generates its proper magnetic field, which tangentially surrounds the black hole horizon and funnels the irreducible to radiation charged particles into two approximately symmetric jets. This self-generated magnetic field provides the outflow channel for the escape of light not irradiated charged matter, which the entropy force rejects from crossing the horizon. Matter accreted by a massive black hole does not arrive at the very horizon as matter, but as charge and massless radiation: mostly photons, possibly~gluons, depending on the strength of the gravitational force (i.e., on the mass $M$ of the black hole whether or not it would be large enough to overcome gluon confinement, which is rather~unbelievable). 

The fraction of radiative energy that hits the black hole and gets trapped consists of photons. These are massless and do not feel the entropy force when encountering the horizon, because the entropy force acts on finite mass particles only. The photons feel, however, the gravitational black hole potential, which is acting on their energy and thus gravitationally deflects their orbits. The photon paths become spirally warped until they ultimately hit the horizon. When this happens, the photons make it across the horizon and enter the interior of the black hole. Looked at from the outside, this takes infinitely long again. It is an open question as to what happens to them inside the horizon, whether~or not they collapse, and whether a singularity forms at all if only photons are available. We do not ponder about those interesting questions here.

The implication is that the mass influx into the hole proceeds via irradiation of matter as {radiative} inflow, not as matter inflow. The black hole is fed by {photons}. Feeding a black hole with small portions of matter in this view proceeds via transformation into radiation. Those parts of matter that do not transform into radiation, protons and the required neutralizing electrons, become expelled along the newly-formed magnetic funnels into space in the form of jets before touching the horizon. The jets either disperse in interaction with distant matter or become part of cosmic radiation. The process of how radiation may tunnel across the horizon is answered by the quantum electrodynamics of this process including the positive entropy potential drop at the horizon the radiation passes when hitting the black hole, which however barely affects the uncharged and massless photons.

Another question concerns the merging of two equal mass massive black holes. This case is of substantial interest because it has been observed in the first detections of gravitational radiation. If~there is just a small mass difference, then probably the two almost equally-strong forces would produce a deformation of the horizons at contact, causing a bubble to evolve, like in the encounter of two soap bubbles. Merging of the horizons takes place at the circumference where the gradients of the entropy become tangential. The holes would start here to merge until the horizon encompasses both holes, with a trapped bubble forming in its common interior and thus becoming invisible to the external observer.

Finally, what happens when asking for the mysterious planckions, Planck particles of mass $M\sim10^{-8}$ kg ($\sim$$10^{19}$ GeV), which may have been created in the Big Bang and are believe to be Planck scale black holes? According to Hawking radiation theory, they should have evaporated in a Planck time of $\sim$$10^{-43}$ s already after production, though it is not clear whether at the Planck scale, one can speak at all of black holes, as inside the planckion, quantum gravity necessarily comes into play, and Hawking's quantum electrodynamical calculations should become invalid. Looked at from the outside, a planckion is its own horizon and thus is fuzzy because its radius and diameter equal spatial uncertainty. Assuming that one still could speak about their surface, entropy, and entropy force, their~entropy would be of the order of $S\sim O(1)$, while the entropy force would remain huge, equal to the Schwarzschild constant, outrunning the gravitational attraction force. Does this mean that planckions would neither radiate, nor be able to merge, keeping one another at a distance and in larger numbers causing some crystal-like texture? In this case, they could have survived (cf., e.g., \citep{carr2016} for contras) since the Big Bang and accumulated in agglomerations like clusters of galaxies where they could well serve (see \citep{garny2016} for pros) as a dark matter candidate. 

\section{Microscopic Phase-Space Density and the Entropy Force}

So far, we just discussed the effect of the entropy force on a particular object: Schwarzschild black holes in astrophysics. We now turn to the general kinetic problem of the microscopic evolution of the particle distribution in many-particle physics. We restrict solely to classical systems, i.e., to~systems that are described on the microscopic level by the classical Liouville equation, respectively its Klimontovich~\citep{klimontovich1967} equivalent in Equation (\ref{eq-7}) {given below, and its hydrodynamic generalization~\citep{gerasimenko2011}.}

Liouville's equation describes the evolution of the microscopic phase space density in $N$-dimensional phase space. On the classical elementary level of indistinguishable point charges, which have some properties like mass $m_a$, possibly some charge $e$ of different sign, and are distributed over a spatial volume $V$ with volume element $d^3q$ and the momentum volume of element $d^3p$ can be described alternatively \citep{klimontovich1967,gerasimenko2011} by an {exact} known phase space density:
\begin{eqnarray}\label{eq-6}
\MC{N}_a^m(\mathbf{p},\mathbf{q},t)\ &=&\ \sum_{i=1}^{N_a} \delta\big(\mathbf{p}-\mathbf{p}_{ai}(t)\big)\delta\big(\mathbf{q}-\mathbf{q}_{ai}(t)\big)\nonumber\\[-1.5ex]
&&\\[-1.5ex]
&\equiv&\ \sum_{i=1}^{N_a} \delta\big(\mathbf{x}-\mathbf{x}_{ai}(t)\big)\nonumber
\end{eqnarray}
which simply counts the number of particles of sort $a$ in the {entire 6$D$}%should it be D? should it be italics? please check the conventions
-phase space volume, such that it is normalized as: 
\begin{equation}
N_a\ = \ \int d^3p\ d^3q\ \MC{N}_a^m(\mathbf{p}, \mathbf{q}, t)
\end{equation}
For a constant particle number, the time dependence is implicit in the particle trajectories $\mathbf{p}_{ai}(t),\mathbf{q}_{ai}(t)$ such that integration has to be performed along all of them. One may note that the microscopic phase space density $\MC{N}_a^m$ is otherwise dimensionless. This is seen from the definition of the delta-functions, which in the integration over phase space simply count numbers, which of course means that the normalization to space and momentum is implicit to them. Later, we will make the normalization more explicit, as this will be required by reference to the entropy.

Since the assumption is that the particles are classical, then in the absence of any particle sources or losses, the particle number in phase space is conserved along all the dynamical trajectories of the particles under their mutual, as well as external forces. In this case, the continuity equation of the particles, i.e., the microscopic Liouville equation in the $N_a$-particle 6d-phase space \citep{klimontovich1967}, reads simply:
\begin{equation}\label{eq-7}
\dot{\MC{N}}_a^m\ \equiv\ \frac{\partial{\MC{N}_a^m}}{\partial t}+\frac{\mathbf{p}}{m_a}\ \mathbf{\cdot}\ \nabla_{\mathbf{q}}\:\MC{N}_a^m+\frac{d\mathbf{p}}{d t}\ \mathbf{\cdot}\ \frac{\partial\MC{N}_a^m}{\partial\mathbf{p}}\ =\ 0
\end{equation}
Of course, here, $\mathbf{\dot{p}}=\mathbf{F}$ is the total force that acts on the particles at their location $\mathbf{q}=\mathbf{q}_{ai}(t)$ and thus on the phase space density, and the two last terms together constitute the Poisson bracket $[\dots]$ in the Liouville equation, which in $N_a$-phase space, the 6d-phase space that within, the $N_a$ particles perform their trajectories, is a tautology.

The entropy force can be compared with other more conventional forces. Let, for simplicity, the~total force $\mathbf{F}=\mathbf{F}_Q+\mathbf{F}_S$ be the sum of the entropy force and of another potential force $\mathbf{F}_Q=-\nabla_qU$, where $U(\mathbf{q},t)$ is the force potential. The entropy force just adds the potential $TS(\mathbf{q},t)$ to the force potential $U$. Note that the entropy-force potential is always {positive}, as already made use of above, because there are neither negative temperatures \citep{dunkel2014} ({{The absence of negative temperatures is immediately clear from the definition of the temperature} 
%footnotes are not allowed according to the journal rule, we moved the footnote here, please confirm.
$T$ as proportional to the mean ensemble averaged square of the momentum fluctuations $T\sim\langle(\delta\mathbf{p})^2\rangle$ of all particles in the volume, which clearly, is a positive definite quantity. Negative temperatures would require imaginary momenta or particle mass. There are no candidates for such particles, though experiments of neutrino oscillations provide negative mean square masses, which, however, are interpreted differently.}), nor are there negative entropies. Clearly, for strong forces acting on the particles and weak entropy gradients, the~entropy force is negligible. This might be the usual case. On the other hand, if on the large scale the inter-particle forces compensate, the entropy force will remain because there is no obvious counterpart that could compensate it. For instance, when dealing with electrostatic interactions only in the absence of any external fields and forces, the microscopic force $\mathbf{F}_Q^m(\mathbf{q},t)=\sum_a e_a\mathbf{\MC{E}}^m(\mathbf{q},t)$ is the Coulomb force acting on the charges $e_a=ae$ with $a=+,-$ in the microscopic electrostatic field $\mathbf{\MC{E}}^m(\mathbf{q},t)$ obeying Maxwell--Poisson's equations:
\begin{equation}\label{eq-8}
\nabla_q\ \mathbf{\cdot}\ \mathbf{\MC{E}}^m=\frac{1}{\epsilon_0}\sum_a\rho_{ae}^m(\mathbf{q},t), \quad \mathbf{\MC{E}}^m=-\nabla_{\mathbf{q}}\Phi^m_e(\mathbf{q},t)
\end{equation}
with electrostatic potential $\Phi^m_e$, and thus, $U=\sum_ae_a\Phi^m_e$. On the microscopic level in phase space, the microscopic electric space charge (not the charge density) of species $a$ is: 
\begin{equation}\label{eq-9}
\rho_{ae}^m(\mathbf{q},t)=e_a\int d^3p\ \MC{N}_a^m(\mathbf{p},\mathbf{q},t), \quad \int d^3q\rho^m_{ae}=\ e_a N_a
\end{equation}
It simply counts all charges in the total volume not relating them to the spatial volume $V_a$ yet. Summing over all species $a$, the total space charge is obtained. On average, it will be zero. The charges are moving, and there is a microscopic current: 
\begin{equation}\label{eq-10}
\mathbf{j}^m_{ae}(\mathbf{q},t)=\frac{e_a}{m_a}\int d^3p\ \mathbf{p}\:\MC{N}_a^m(\mathbf{p},\mathbf{q},t)\ = \ e_a \mathbf{v}_a(\mathbf{q},t) N_a
\end{equation}
with $\mathbf{v}_a(\mathbf{q},t)$ the average velocity of particles in group $a$. It gives rise to an internal magnetic field, which, in the electrostatic approximation, is relativistically small and is thus neglected (e.g., \citep{klimontovich1967}), though this is not completely correct, because in a linear theory of fluctuations, it should be taken into~account. 

Electrostatic interactions have been the subject of exhaustive investigations in the literature. Here,~they serve only as another force field against which the entropy force can be compared. The~striking difference is that for the entropy force, no field is generated because there is no entropy charge comparable to $e_a$ and, hence, no singularity that would act as the source of the entropy field. In other words, the entropy field is, in contrast to the electric field, not related to field equations and thus lacks a field theory. Disorder lacks any elementary source not being a field, at least in classical physics. 

The entropy of a system entering the first law of total energy conservation is an integral quantity. In order to refer to it on the elementary level of the microscopic kinetic equation in 6d-phase space, one has to return to its microscopic definition as the phase space average of the probability distribution. 

It is convenient to define an entropy phase space density by referring to Gibbs--Boltzmann's definition of entropy through the probability density. Entropy density will then be obtained by integrating out the momentum space coordinates in the usual way. In the definition of the phase space density of entropy, we will at this point not yet make the assumption that the phase space volume is constant, but include the spatial dependence as well. This is advantageous because it allows making use of phase space densities. Integrating out the volume can be done at a later stage. With this in mind, the ``{microscopic Boltzmann entropy phase-space density}'' of species $a$ becomes ({{A number of other definitions or generalizations of entropy different from Boltzmann--Gibbs have been put forward in the near past} %move the footnote here, please confirm.
 \citep{renyi1970,wehrl1978,tsallis1988,treumann1999,treumann2004,treumann2008,treumann2014}, the physical, not the statistical meaning of which is not entirely clear. Though~the theory could be extended to include those, we will neither refer to, nor use them in this~note.}):
\begin{equation}\label{eq-11}
\MC{S}_{aB}^m(\mathbf{p},\mathbf{q},t) = -\log\:\MC{N}_a'^m(\mathbf{p},\mathbf{q},t), \quad \MC{N}_a'^m=\MC{N}_a^m/N_a
\end{equation}
where the phase space density has been normalized to the total number $N_a$ of particles of species $a$. This makes the argument of the logarithm smaller than one, of which the negative sign takes care. A~definition like this leans on Boltzmann's proposal. It is incomplete on the microscopic level because the entropy is a {collective} quantity, which is obtained by integrating over momentum space with the microscopic phase space density $\MC{N}_a$ as the weight. One thus has as {microscopic $N$-particle entropy phase-space density} in $(\mathbf{p},\mathbf{q})$-space: 
\begin{equation}\label{eq-12}
\MC{S}_a^m(\mathbf{p},\mathbf{q},t) = -\MC{N}_a^m(\mathbf{p},\mathbf{q},t)\: \log\:\MC{N}_a'^m(\mathbf{p},\mathbf{q},t)
\end{equation} 
This microscopic phase-space density of the entropy is not the entropy itself, which is a function solely of the space coordinates. The $N_a$-particle entropy of sort $a$ is explicitly obtained by the integration of~(\ref{eq-12}) over the entire momentum phase space:
\begin{eqnarray}\label{eq-14}
S_a^m(\mathbf{q},t)&=&\int d^3p\ \MC{S}_a^m(\mathbf{q},\mathbf{p},t)\nonumber\\[-1.5ex]
&&\\[-1.5ex]
&\equiv&\int d^3p'd^3q'\delta(\mathbf{q}-\mathbf{q}')\ \MC{S}_a^m(\mathbf{q}',\,\mathbf{p}',\,t)\nonumber
\end{eqnarray}
and is always positive, as is easily seen from the definition of the microscopic phase space density, a positive quantity, and the above choice of entropy. Summing over all particle species $a$ then gives the total entropy. Moreover, because information is transported via some field, for instance the electromagnetic field, the time under the integral in Equation (\ref{eq-12}) is the retarded time $t^R=t-|\mathbf{q}-\mathbf{q}'|/c$ where $c$ is the velocity of signal/information transport between the particles at locations $\mathbf{q}$ and $\mathbf{q}'$ in the real-space subspace of the 6d-phase space \citep{wheeler1945,wheeler1949}. In conventional kinetic theory, retardation is neglected because $c$ is the velocity of light, and the distances between particles are usually less than $ct$. This is also assumed in the following.

There is a direct correspondence between this real space microscopic entropy density and the real space charge density $\rho^m(\mathbf{q},t)$. Both enter the force term via taking the spatial gradient. The difference is that for the electric charge density, this step passes through the electric field $\mathbf{\MC{E}}^m$, which is generated by the space charges. Repeated again, for the entropy, there is no such field, nor field equation in classical physics. Entropy is not a charge of some entity and thus does not generate a field. Taking its spatial gradient directly provides the force that acts on the particle at location $\mathbf{q}$.

\section{Kinetic Equation with Entropy Force}
The entropy force acting on species $a$ is the negative gradient of Equation (\ref{eq-14}). This leads to a repulsive force, independent of any charge. It adds to the potential $U$ in Klimontovich's equation. Thus,~taking it into account in Equation (\ref{eq-7}), it becomes clear that it does not affect the particle number and thus does not imply any important change in the microscopic $N_a$-particle phase space density $\MC{N}_a^m$. The main interest is in its effect on the one-particle kinetic phase space distribution function $f_a(\mathbf{x},t)$. This is defined through the ensemble-averaged $N_a$-particle phase space density:
\begin{equation}\label{eq-15}
\frac{N_a}{V_a}f_{a}(\mathbf{x},t)=\big\langle\MC{N}_a^m(\mathbf{x},t)\big\rangle
\end{equation}
where $\langle\dots\rangle$ indicates the ensemble average, and explicitly for the one-particle distribution:
\begin{eqnarray}\label{eq-16}
f_a(\mathbf{x}_{a1},t)&=&V_a\int f_Nd^6\mathbf{x}_{a2}\dots d^6\mathbf{x}_{aN_a}
\prod_{b\neq a}d^6\mathbf{x}_{b1}\dots d^6\mathbf{x}_{bN_b}\\
\big\langle\MC{N}^m_a(\mathbf{x},t)\big\rangle&=&N_a\int\delta(\mathbf{x}-\mathbf{x}_{a1})f_N\prod_ad^6\mathbf{x}_{a1}\dots d^6\mathbf{x}_{aN_a}
\end{eqnarray}
with $V_a\equiv V$ the spatial volume occupied by the indistinguishable particle sort $a$. $f_N$ is the $N$-particle distribution function, and the integration is with respect to all indistinguishable particles $N-1$, but one, the particle with coordinates $\mathbf{x}_{a1}$, as has been defined by \citet{klimontovich1967}. In fact, the distribution function $f_N$ is not explicitly given. It can be resolved on the way of sequentially stepping up the ladder from the one-particle distribution function to higher order distribution functions, which depend on one, two, three, or more indistinguishable particles.

Before proceeding to rewriting the Klimontovich Equation (\ref{eq-7}), it is necessary to investigate what happens to the entropy when performing the ensemble average implied in the former equations. We~do actually not need the entropy itself, rather its spatial gradient, i.e., we need the spatial gradient of the entropy-phase space density (\ref{eq-12}). For this, we have:

\begin{equation}\label{eq-18}
-\nabla_{\mathbf{q}}\MC{S}_a^m =\Big(1+\log\:\MC{N}'^m_a\Big)\nabla_{\!\!\mathbf{q}}\:\MC{N}_a^m
\end{equation}
The entropy force is the integral of the gradient of Equation (\ref{eq-14}) over the primed phase space:
\begin{eqnarray}
\mathbf{F}_{S}^{\,a}(\mathbf{q},t)\ &=&T\nabla_{\mathbf{q}}\int d^3p'd^3q'\delta(\mathbf{q}-\mathbf{q}')\times\nonumber\\[-2ex]
&&\\[-1ex]
~~&\times&\MC{N}_a^m(\mathbf{p}',\,\mathbf{q}',\,t)\: \log\:\MC{N}_a'^m(\mathbf{p}',\,\mathbf{q}',\,t)\nonumber
\end{eqnarray}
Reduction of the entropy-force term in Klimontovich's equation requires performing the ensemble average of the term:
\begin{equation}
\mathbf{F}_{S}^{\,a}(\mathbf{q},t)\ \mathbf{\cdot}\ \frac{\partial}{\partial\mathbf{p}}\MC{N}_a^m(\mathbf{q},\mathbf{p},t)
\end{equation}
In order to do so, we need to consider different groups of particles such that: 
\begin{eqnarray}
\mathbf{F}_S(\mathbf{q},t)\ &=&T\sum_b\nabla_{\mathbf{q}}\int d^3p'd^3q'\delta(\mathbf{q}-\mathbf{q}')\times\nonumber\\[-2ex]
&&\\[-1ex]
~~&\times&\MC{N}_b^m(\mathbf{p}',\,\mathbf{q}',\,t)\: \log\:\MC{N}_b'^m(\mathbf{p}',\,\mathbf{q}',\,t)\nonumber
\end{eqnarray}
This produces formally the entropy force contribution to the Klimontovich equation:
\begin{eqnarray}
\mathbf{F}_S&\mathbf{\cdot}&\frac{\partial}{\partial\mathbf{p}}\MC{N}_a^m =T\nabla_{\mathbf{q}}\sum_b\int d^3p'd^3q'\delta(\mathbf{q}-\mathbf{q}')\ \mathbf{\cdot}\nonumber\\[-2ex]
&&\\[-1ex]
&\mathbf{\cdot}&\frac{\partial}{\partial\mathbf{p}}\bigg\langle\MC{N}_a^m(\mathbf{p},\mathbf{q},t)\MC{N}_b^m(\mathbf{p}',\,\mathbf{q}',\,t)\: \log\:\MC{N}_b'^m(\mathbf{p}',\,\mathbf{q}',\,t)\bigg\rangle\nonumber
\end{eqnarray}
{where $\langle\dots\rangle$ indicates the ensemble average, and we have used Equation (\ref{eq-18}). The momentum differentiation affects only terms containing the phase space density. This leads to the appearance of the logarithmic term on the right and introduces a third-order correlation term.} The $(N-1)$-particle ensemble-averaged term provides problems because it contains the logarithm of the phase space density. In a somewhat severe approximation, we may assume that the logarithm is a slowly-varying function. Its argument is smaller than one such that it can be expanded, which {yields:} %please check if there is a extra $\approx$ in equation (27).
\begin{eqnarray}\label{eq-23}
&&\bigg\langle\MC{N}_a^m(\mathbf{p},\mathbf{q},t)\:\MC{N}_b^m(\mathbf{p}',\,\mathbf{q}',\,t)\: \log\:\MC{N}_b'^m(\mathbf{p}',\,\mathbf{q}',\,t)\bigg\rangle \nonumber\\[-1.7ex]
&&\\[-1.7ex]
&\approx&\bigg\langle\MC{N}_a^m(\mathbf{p},\mathbf{q},t)\:\MC{N}_b^m(\mathbf{p}',\,\mathbf{q}',\,t)\: \Big(\MC{N}_b'^m(\mathbf{p}',\,\mathbf{q}',\,t)-1\Big)\bigg\rangle\nonumber
\end{eqnarray}
This generates the ensemble-averaged Klimontovich {equation:}%please check if there is a extra $\cdot$ in equation (28).
\begin{eqnarray}\label{eq-24}
\frac{\partial\langle\:\MC{N}_a^m\rangle}{\partial t}&+&\frac{\mathbf{p}}{m_a}\ \mathbf{\cdot}\ \nabla_q\langle\:\MC{N}_a^m\rangle-T\nabla_q\sum_{b\neq a}\int d^6x'\ \delta(\mathbf{q}-\mathbf{q}')\ \nonumber\\[-2ex]
&&\\[-1ex]
&\mathbf{\cdot}&\frac{\partial}{\partial\mathbf{p}}\Big\langle\MC{N}_a^m(\mathbf{x},t)\:\MC{N}_b^m(\mathbf{x}',\,t)\Big\rangle= -\Big\langle\MC{C}_a^S(\mathbf{x},t)\Big\rangle\nonumber
\end{eqnarray}
The average purely entropic collision term on the right collects the third-order {correlations:}%please check if there is a extra $\cdot$ in the following equation.
\begin{eqnarray}
\Big\langle\MC{C}_a^S(\mathbf{x},t)\Big\rangle &=&T\nabla_q\sum_{b\neq a}\int d^6x'\ \delta(\mathbf{q}-\mathbf{q}')\ \nonumber\\
&\mathbf{\cdot}&\frac{\partial}{\partial\mathbf{p}}\Big\langle\MC{N}_a^m(\mathbf{x},t)\:\MC{N}_b^m(\mathbf{x}',\,t)\:\MC{N}_b'^m(\mathbf{x}',\,t)\Big\rangle\nonumber
\end{eqnarray}
 In these expressions, we have, for simplicity of writing, only included the entropy force term. One~trivially adds the microscopic Coulomb or any other force term to this if required (Note, however, that adding the microscopic gravitational force causes problems because it remains uncompensated ({{As in any kinetic theory, this is an important difference between gravitation and any other force. It~implies that in kinetic theory gravitation can only be included consistently in a general relativistic quantum gravitation where gravitation is balanced by quantum fluctuations.}}).). %move the footnote here, please confirm.
The entropy force term resembles the latter, but lacks a charge singularity. This is replaced by the spatial derivative of the delta-function, which appears under the integral. 

The main difference is that already on this very basic level, the presence of the entropy force contributes a purely entropic dissipative term $\langle\MC{C}_a^S(\mathbf{x},t)\rangle$, which has been transferred to the right in the above expression. This term arises due to the generation of entropy in the system. It is a three-particle correlation term, as will become clear below. It is caused by the logarithm in the entropy, the continuous growth of entropy in a many-particle system. Whether it can be neglected as being of higher order is a subtle question. It causes collisionless dissipation in the presence of entropy. Since this effect is non-collisional, when neglecting particle collisions, one must take care whether its neglect is allowed. Below, we show that, however, dissipation is proportional to the inverse particle number $N_a^{-1}$ and can in most cases for very large numbers of particles be neglected.

The next step in this theory is to relate the last equation to the one-particle distribution function defined in Equation (\ref{eq-15}). Following \citet{klimontovich1967}, this is achieved via considering the fluctuations:
\begin{equation}\label{eq-25}
\delta\MC{N}_a^m(\mathbf{x},t)=\MC{N}_a^m(\mathbf{x},t)-\Big\langle\MC{N}_a^m(\mathbf{x},t)\Big\rangle
\end{equation}
When ensemble-averaged, these deviations from the mean phase-space density vanish, and we {have:}%please check if there is a extra + in equation (30).
\begin{eqnarray}\label{eq-26}
\Big\langle\MC{N}_a^m(\mathbf{x},t)\MC{N}_b^m(\mathbf{x}',\,t)\Big\rangle&=&\Big\langle\MC{N}_a^m(\mathbf{x},t)\Big\rangle\Big\langle\MC{N}_b^m(\mathbf{x}',\,t)\Big\rangle\nonumber\\[-2ex]
&&\\[-2ex]
&+&\Big\langle\delta\MC{N}_a^m(\mathbf{x},t)\:\delta\MC{N}_b^m(\mathbf{x}',\,t)\Big\rangle\nonumber
\end{eqnarray}
We can now make use of the definition of the one-particle distribution function $f_a(\mathbf{x},t)$ by~\mbox{\citet{klimontovich1967}}. Define the particle density $n_a=N_a/V_a$ to {obtain:} %please check if there is a extra + in equation (31).
\begin{eqnarray}\label{eq-27}
\Big\langle\MC{N}_a^m(\mathbf{x},t)\MC{N}_b^m(\mathbf{x}',\,t)\Big\rangle&=& n_an_b\Big[f_a(\mathbf{x},t)f_b(\mathbf{x}',\,t)\nonumber\\[-2ex]
&&\\[-1ex]
+g_{ab}(\mathbf{x},\mathbf{x}',\,t)\Big]
&+&\delta_{ab}\delta(\mathbf{x}-\mathbf{x}')n_af_a(\mathbf{x},t)\nonumber
\end{eqnarray}
Here, $g_{ab}(\mathbf{x},\mathbf{x}',\,t)$ is the two-particle correlation function, which results from the ensemble-averaged product of the fluctuations $\delta\MC{N}^m$ of the phase space density in the last term on the right in Equation~(\ref{eq-26}). The three terms in the expression (\ref{eq-27}) are of the same structure as in the ordinary one-particle kinetic theory \citep{klimontovich1967}. One may note that the last term, which is linear in the distribution function, simply becomes absorbed in the convective term in the kinetic equation. In non-relativistic, theory it just causes a translation. We can immediately write down the one-particle kinetic equation including the entropy force. One must, however, take care of to which terms the gradient and momentum operations apply. This yields the {result:}%please check if there is a extra $\times$ in equation (32).

\begin{eqnarray}\label{eq-28}
&&\frac{\partial f_a}{\partial t}\:+\:\frac{\mathbf{p}}{m_a}\ \mathbf{\cdot}\ \nabla_{\mathbf{q}}f_a -T\:\nabla_{\mathbf{q}}^b\sum_bn_b\int d^6x'\nonumber\\[-2ex]
&&\\[-1ex]
&&\times\:\delta(\mathbf{q}-\mathbf{q}')\ \mathbf{\cdot}\ \frac{\partial}{\partial\mathbf{p}}\Big[f_a(\mathbf{x},t)f_b(\mathbf{x}',\,t)\Big]= \Big(\MC{G}_{ab}^{S(\mathrm{x},t)}-\Big\langle\MC{C}_a^{S(\mathrm{x},t)}\Big\rangle\Big)(\mathbf{x},t)\nonumber
\end{eqnarray}
which, when integrating over the primed spatial coordinate, simplifies {to:}%please check if there is a extra $\cdot$ in equation (33).
\begin{eqnarray}\label{eq-29}
&&\frac{\partial f_a}{\partial t}\:+\:\frac{\mathbf{p}}{m_a}\ \mathbf{\cdot}\ \nabla_{\mathbf{q}}f_a -T\:\frac{\partial}{\partial\mathbf{p}}f_a(\mathbf{q},\mathbf{p},t) \nonumber\\[-2ex]
&&\\[-1ex]
&&\mathbf{\cdot}\nabla_{\mathbf{q}}\ \sum_bn_b\int d^3p'\ f_b(\mathbf{q},\mathbf{p}',\,t)= \Big(\MC{G}_{ab}^{S(\mathrm{x},t)}-\Big\langle\MC{C}_a^{S(\mathrm{x},t)}\Big\rangle\Big)(\mathbf{x},t)\nonumber
\end{eqnarray}
In this expression on the right, the term $\MC{G}_{ab}$ results from the two-particle correlation term $g_{ab}(\mathbf{x},\mathbf{x}',\,t)$. It corresponds to what in kinetic theory is understood as direct particle collisions. The last expression contains the integral over the primed momentum space. Only the distribution $f_b$ depends on this integration. It is therefore convenient to define the number density $\rho_a$ of species $a$ as:
\begin{equation}
\rho_a(\mathbf{q},t)= n_a\int d^3p\ f_a(\mathbf{q},\mathbf{p},t)
\end{equation}
{and the last expression just includes the global entropy force {term:}}%please check if there is a extra $=$ in equation (35).
\begin{eqnarray}\label{eq-29a}
\frac{\partial f_a}{\partial t}\:&+&\:\frac{\mathbf{p}}{m_a}\ \mathbf{\cdot}\ \nabla_{\mathbf{q}}f_a -\ T\:\bigg( \nabla_{\mathbf{q}}\ \sum_b\rho_b(\mathbf{q},t)\bigg)\ \mathbf{\cdot}\ \frac{\partial f_a}{\partial\mathbf{p}} \nonumber\\[-1ex]
&&\\[-1ex]
&= &\Big(\MC{G}_{ab}^{S(\mathrm{x},t)}-\Big\langle\MC{C}_a^{S(\mathrm{x},t)}\Big\rangle\Big)(\mathbf{x},t)\nonumber
\end{eqnarray}
The sum is over all particle components, implying the total number density. Thus, the entropy force term simply adds to any other potential force term in the kinetic equation. This is true already to first order in the expansion of the logarithmic term in the definition of the entropy. In the case of charged particles, such a force is the Coulomb force or the Lorentz force, when including magnetic fields. The difference is, however, that this force term does not depend on charge while acting on the microscopic particle phase space distribution. It resembles the gravitational force, but does not contain its inverse square dependence on the inter-particle distance. This is advantageous as it releases from the necessity of compensation. On the other hand, the new force term introduces another non-linearity contained in the density, which itself is the integral of the distribution function. 

The entropy force resembles a pressure force on the kinetic level. With zero right-hand side in the kinetic equation, it conserves particle number. This is a rather simple result, which, of course, could~have been anticipated, without reference to any complicated derivation from first principles as done here, by adding a macroscopic entropy force to the force terms. 
 
The ensemble-averaged term $\big\langle\MC{C}_a^S\big\rangle$ is a purely entropic lowest order $\big($in the smallness of $\MC{N}'^m_a)$ dissipation term for which, in conventional kinetic theory, no equivalence arises. This term is, however,~small and thus negligible, as will be shown in the next section.

In the collisionless kinetic theory of forces between particles, any non-collisional dissipation term caused by particle interactions via their fields yields correlations, which can be neglected, respectively discussed away by comparing dissipation and collisionless scales. The entropic dissipation term instead remains because it is not caused by particle collisions, nor wave--particle interactions. There is no entropy source field that leads to the correlations between particles. Rather, it is the inhomogeneity in the macroscopic disorder that is responsible for the fluctuations and the appearance of the dissipative entropic correlations between the fluctuations leading to the dissipation term. Hence, this term remains even under completely collisionless conditions. Once disorder exhibits spatial structure, it will always be present. In the next section, we provide the explicit versions of these two terms.

\section{Dissipative Terms}
In order to complete the theory, one needs to express the two dissipative terms in the final kinetic Equation (\ref{eq-28}). The collision term $\MC{G}_{ab}$ is of the same structure as the Coulomb collision term \citep{klimontovich1967,treumann2017}. It~adds to the latter: 
\begin{equation}\label{eq-30}
\MC{G}_{ab}^S= T\sum_bn_b\nabla_{\mathbf{q}}\:\mathbf{\cdot}\:\int d^6x'\delta(\mathbf{q}-\mathbf{q}')\frac{\partial}{\partial\mathbf{p}}g_{ab}(\mathbf{x},\mathbf{x}',\,t)
\end{equation}
{In the particular case that the correlation $g_{ab}$ does not contain any singularity, this expression has no singularity at $\mathbf{q}=\mathbf{q}'$} other than that in the derivative of the delta-function, which replaces $\mathbf{q}'\to\mathbf{q}$ in the correlation function $g_{ab}$ when the integration is carried out. The remaining expression becomes:
\begin{equation}\label{eq-31}
\MC{G}_{ab}^S(\mathbf{q},\mathbf{p},t)= T\:\frac{\partial}{\partial\mathbf{p}}\ \mathbf{\cdot}\ \sum_bn_b\int d^3p'\ \nabla_{\mathbf{q}}g_{ab}(\mathbf{q},\mathbf{p},\mathbf{p}',\,t)
\end{equation} 
The presence of the entropy force thus contributes to the collision term via the particle correlation function. Clearly, due to the strong Coulomb force at short distances, the particle interaction on the short scales is dominated by the Coulomb force. However, at distances larger than the Coulomb collision length, the entropic collisional interaction remains. In a charge collisionless plasma, for~instance, the~Coulomb term causes charge screening felt inside the Debye sphere and eliminates the microscopic electric field between the charges on scales larger than the Debye length $\lambda_D$. On such scales, entropic dissipation might enter the scene. Since it is proportional to temperature $T$ and also number density $\rho$, it has the character of a collisional contribution of the pressure in the inhomogeneities of the entropy.

One may take notice that the spatial gradient operator can be taken out of the integral in the last expression. This allows writing:
\begin{equation}
\MC{G}^S_{ab}(\mathbf{q},\mathbf{p},t)=T\:\nabla_{\mathbf{q}}\ \mathbf{\cdot}\ \frac{\partial}{\partial\mathbf{p}} G_a^S(\mathbf{q},\mathbf{p},t)
\end{equation}
where we introduced the entropic correlation integral:
\begin{equation}\label{G}
G_a^S(\mathbf{q},\mathbf{p},t) = \sum_bn_b\int d^3p'\ g_{ab}(\mathbf{q},\mathbf{p},\mathbf{p}',\,t)
\end{equation}
We will return to the discussion of this correlation integral below, because it contains the most interesting effect introduced by the entropy force. 

{In general, the correlation will contain contributions from the forces that depend on local charges. This leads from the field equations to singularities in the correlation function, as the example of the Coulomb force suggests. The simplified form of the correlation term in the dissipation function retains the form in Equation (\ref{eq-30}) and must be explicitly spelled out. Exchanging the derivatives, this can be~written as:
\begin{equation}
\MC{G}_{ab}^S= T\frac{\partial}{\partial\mathbf{p}}\:\mathbf{\cdot}\:\sum_bn_b\nabla_{\mathbf{q}}\int d^6x'\delta(\mathbf{q}-\mathbf{q}')g_{ab}(\mathbf{x},\mathbf{x}',\,t)
\end{equation}
}

Let us now turn to the remaining term $\big\langle\MC{C}_a^S\big\rangle$. From Equations (\ref{eq-23}) and (\ref{eq-24}), one realizes that it contains the product of three microscopic phase space densities before taking the ensemble average. This complicates its calculation. In analogy to Equations (\ref{eq-26}) and (\ref{eq-27}), it requires the introduction of higher order correlations. Formally, this is quite simple as it has been pioneered by \citet{klimontovich1967} how one would have to deal with it in this case. We need the third-order ensemble average, which becomes in the same way as (\ref{eq-27}):
\begin{eqnarray}\label{eq-32}
\Big\langle\MC{N}_a^m\MC{N}_b^m\MC{N}_c^m\Big\rangle = n_an_bn_cf_{abc}(\mathbf{x},\mathbf{x}',\,\mathbf{x}'') &&\nonumber\\[-1ex]
+\delta_{ab}n_an_c\delta(\mathbf{x}-\mathbf{x}')f_{ac}(\mathbf{x},\mathbf{x}'')&&\nonumber\\
+\delta_{ac}n_an_b\delta(\mathbf{x}-\mathbf{x}'')f_{ab}(\mathbf{x},\mathbf{x}')&&\\
+\delta_{bc}n_an_c\delta(\mathbf{x}'-\mathbf{x}'')f_{ac}(\mathbf{x},\mathbf{x}'')&&\nonumber\\
+\delta_{ab}\delta_{bc}\delta(\mathbf{x}-\mathbf{x}')\delta(\mathbf{x}-\mathbf{x}'')f_{a}(\mathbf{x})&&\nonumber
\end{eqnarray}
For convenience, we dropped the common variable $t$. The two- and three-particle distribution functions $f_{ab},\ f_{abc}$ {read:}%please check if there are two extra $+$ in equation (42).
\begin{eqnarray}\label{eq-33}
f_{ab}(\mathbf{x},\mathbf{x}')&=&f_a(\mathbf{x})f_b(\mathbf{x}')+g_{ab}(\mathbf{x},\mathbf{x}')\nonumber\\
f_{abc}(\mathbf{x},\mathbf{x}',\,\mathbf{x}'')&=&f_a(\mathbf{x})f_b(\mathbf{x}')f_c(\mathbf{x}'')\nonumber\\[-2.5ex]
&&\\[-1ex]
&+&f_a(\mathbf{x})g_{bc}(\mathbf{x}',\,\mathbf{x}'') +f_b(\mathbf{x}')g_{ac}(\mathbf{x},\mathbf{x}'')\nonumber\\ 
&+&f_c(\mathbf{x}'')g_{ab}(\mathbf{x},\mathbf{x}')+g_{abc}(\mathbf{x},\mathbf{x}',\,\mathbf{x}'')\nonumber
\end{eqnarray}
With their help, the dissipation function can be constructed. However, its structure simplifies substantially because in our case, on the left in the first line in Equation (\ref{eq-32}), the microscopic phase space densities of index $b$ and $c$ are identical, and only the $f_{ab}$ contributes. Hence, the ensemble average~becomes:
\begin{equation}\label{eq-34}
\Big\langle\MC{N}_a^m\MC{N}_b^m\MC{N}'^m_c\delta_{bc}\delta(\mathbf{x}'-\mathbf{x}'')\Big\rangle = n_an_bf_{ab}(\mathbf{x},\mathbf{x}')\frac{1}{N_b}
\end{equation}
Therefore, only the two-particle distribution function $f_{ab}$ would be relevant in the determination of the dissipative term, i.e., in the first line in Equation (\ref{eq-33}). This is, however, the same as what we already used in Equation (\ref{eq-27}). To this result, one has to apply the operation of space and momentum differentiation. Hence, the collisionless dissipative term contributed by the entropy force is of the same kind as the collision term $\MC{G}_{ab}^S$ it contributes, though being of a different sign. In other words, the two terms would cancel to first order if there were not the normalization to particle number $N_b$. Since~$N_b\approx N_a\gg 1$, we find that the collisionless dissipation due to the entropy force $\langle\MC{C}^S_a\rangle\ll\MC{G}^S_{ab}$ is small and can be neglected in comparison with the entropic collision term. 

Thus, it is the collisional correlation term $\MC{G}^S_{ab}(\mathbf{x},t)$ (\ref{eq-31}) that is retained as a long-range collisional dissipation introduced by the presence of the entropy force. It adds to the Coulomb collisions and might become important on the large scales much larger than the Debye scale or any other inter-particle interaction scale. this important result suggests that the entropic dissipation is a mesoscale, respectively macro effect. We should, however, point out here that we are still dealing with a non-relativistic theory. At large scales, transport and propagation of information, respectively entropy, cannot be neglected anymore, and the theory has to given a covariant relativistic formulation. 

\section{Kinetic Equation for Fluctuations}
The remaining problem is the behavior of fluctuations. These are defined as deviations in the one-particle phase space distribution $f_a$ from its mean ``equilibrium'' value $\bar{f}_a$ as:
\begin{equation}\label{eq-35}
\delta f_a=f_a-\bar{f}_a, \qquad \overline{\delta f}_a=0 
\end{equation}
The evolution equation of the fluctuations is obtained from Equation (\ref{eq-29}) via subtracting the averaged kinetic equation:

\begin{eqnarray}\label{eq-36}
&&\frac{\partial \bar{f}_a}{\partial t}\:+\:\frac{\mathbf{p}}{m_a}\ \mathbf{\cdot}\ \nabla_{\mathbf{q}}\bar{f}_a -T\:\sum_bn_b\int d^3p'\times\nonumber\\[-2ex]
&&\\[-1.5ex]
&&\times\ \nabla_{\mathbf{q}}^b\ \mathbf{\cdot}\ \frac{\partial}{\partial\mathbf{p}}\Big[\bar{f}_a(\mathbf{q},\mathbf{p},t)\bar{f}_b(\mathbf{q},\mathbf{p}',\,t)+ \overline{\delta f_a\delta f_b}\Big]\: =\ \overline{\MC{G}_{ab}^S}\nonumber
\end{eqnarray}
The mean collision term on the right contains all the contributions of the correlations of the mean and fluctuating quantities. Being interested only in linear fluctuations and assuming that the collisions are weak enough to not contribute to the evolution of fluctuations, we drop this term in the following. Subtracting from the complete kinetic equation, the fluctuations obey the non-collisional {equation:}%please check if there is a extra $\times$ in equation (46).
\begin{eqnarray}\label{eq-37}
&&\frac{\partial \delta{f}_a}{\partial t}\:+\:\frac{\mathbf{p}}{m_a}\: \mathbf{\cdot}\: \nabla_{\mathbf{q}}\,\delta{f}_a -T\:\sum_bn_b\int d^3p' \nonumber\\[-1.25ex]
&&\times\ \nabla_{\mathbf{q}}^b\ \mathbf{\cdot}\ \frac{\partial}{\partial\mathbf{p}}\Big[\delta{f}_a(\mathbf{q},\mathbf{p},t)\bar{f}_b(\mathbf{q},\mathbf{p}',\,t)\Big]= \\[-0.5ex]
&&T\:\sum_bn_b\int d^3p'\nabla_{\mathbf{q}}^b\ \mathbf{\cdot}\ \frac{\partial}{\partial\mathbf{p}}\Big[\bar{f}_a(\mathbf{q},\mathbf{p},t)\delta{f}_b(\mathbf{q},\mathbf{p}',\,t)- \overline{\delta f_a\delta f_b}\Big]\nonumber
\end{eqnarray}
This expression still contains the average $\overline{\delta f_a\delta f_b}$ of the squared fluctuations. If this is a constant on the fluctuation time scale, then the equation can be rescaled. In linear theory, it would be neglected to first order and taken into account to second order in a quasi-linear approach. 

Again, carrying out the integration with respect to $\mathbf{p}'$, the last expression simplifies to:
\begin{eqnarray}\label{eq-38}
&&\frac{\partial \delta{f}_a}{\partial t}\:+\:\frac{\mathbf{p}}{m_a}\mathbf{\cdot}\: \nabla_{\mathbf{q}}\,\delta{f}_a -T\ \frac{\partial\delta{f}_a}{\partial\mathbf{p}}\: \mathbf{\cdot}\ \nabla_{\mathbf{q}}\bigg(\sum_b{\bar{\rho}_b(\mathbf{q},t)}\bigg)\nonumber\\[-1.25ex]
&&\\[-0.5ex]
&&=T\:\sum_{b\neq a}n_b\int d^3p'\nabla_{\mathbf{q}}^b\ \mathbf{\cdot}\ \frac{\partial}{\partial\mathbf{p}}\Big[\bar{f}_a(\mathbf{x},t)\delta{f}_b(\mathbf{q},\mathbf{p}',\,t)- \overline{\delta f_a\delta f_b}\Big]\nonumber
\end{eqnarray}
where in the term on the right-hand side, we retained the fluctuation in the distribution function, not~replacing it with the density fluctuation $\delta\rho_b(\mathbf{q},t)$ for the obvious reason that this is an equation for the fluctuations in the distribution function itself. The term on the right couples all fluctuations in the different particle components $b\neq a$ to $\bar{f}_a$.

The linear theory is still complicated by the fact that it contains the sum over the particle correlations. Here, one must include all particles contained in the medium. Moreover, we have written here only those terms that result from the inclusion of the entropy force. To these terms, one must add the electromagnetic force terms. Since the electromagnetic and entropy forces superimpose, this~does not produce any additional mixing, but simply adds those common and well-known terms that take care of the electromagnetic interactions. In this sense, the formal theory is complete. The ranges of the two different forces acting on the particle populations are vastly different because the entropy force is a collective force, which does not originate from any elementary charge. There is no singularity of the entropy that could give rise to an entropy field. The search for such singularities is outside classical~physics.

\section{Evolution of Entropic Phase Space Density}
Having defined the entropic phase space density in Equation (\ref{eq-12},) the question arises how it possibly evolves in phase space. Since the entropy is given as the phase space integral with respect to the entropic phase space density, an always positive quantity, this question is not senseless. Once~knowing its evolution, the entropy can be calculated by integration. Moreover, if an equation for the phase space density can be obtained, its entropic momentum should yield an evolution equation for the entropy, which, essentially, in the long-term limit should be the fundamental thermodynamic laws, while in the short term, it should give the evolution equation of entropy with time. In order to construct the entropic phase space equation, we multiply Equation (\ref{eq-7}) by $-\log\MC{N}'^m_a$ to obtain:
\begin{equation}
\partial_t\MC{S}_a^m+\Big[\MC{H}_{N_a},\MC{S}_a^m\Big]-
\Big\{\partial_t\MC{N}_a^m+\Big[\MC{H}_{N_a},\MC{N}_a^m\Big]\Big\}=0
\end{equation}
The term in the braces vanishes identically, yielding ultimately:
\begin{equation}
\partial_t\MC{S}_a^m+\Big[\MC{H}_{N_a},\MC{S}_a^m\Big]=0
\end{equation}

We thus find the almost trivial result that the microscopic entropic phase-space density $\MC{S}^m_a$ itself satisfies the Liouville equation, i.e., the continuity equation for the entropic phase space density in the phase space. It thus evolves in the microscopic particle phase space like a dissipationless fluid. This is just another expression for the fact that (classically), there are no microscopic sources of entropy, nor is there any entropic field. Entropy is just disorder in the particles. 

However, the microscopic entropic density in phase space is not entirely independent of any disorder. It acts back on itself via the integral entropy force term contained in the above Hamiltonian $\MC{H}_{N_a}$. It provides an entropy potential contribution $U_S=TS$ to the Hamiltonian with $S$, the momentum space integral of the entropy phase space density $\MC{S}^m_a\{\MC{N}^m_a\}$, {which itself is a function} %please check that the correction did not change your meaning
 of the phase space density $\MC{N}^m_a$. Though the structure of the kinetic equation for the entropy density $\MC{S}_a^m$ remains the same as that of the phase space density $\MC{N}_a^m$, both containing the entropy force term and becoming integro-differential equations, the phase space density is determined by the integral entropy density through the entropy force. This force is obtained by adding up all contributions over all phase space. This shows that both the kinetic equation for the particle density and the kinetic equation for the entropy density in phase space must be solved together as both are intimately related. 

One can interpret this result in the way that the {holographic reaction} of the integral entropy on the evolution of the phase space density of the entropy {appears like an elementary entropy source}. This is not unsatisfactory, because it can hardly be expected that a microscopic source of entropy would exist as there are no entropy charges and no entropy fields in the world, at least not classically. Instead, the elementary source arises from the non-linear self-interaction of the entropy. This is a rather important conclusion in that, presumably, little will be changed when including quantum effects in a quantum mechanical treatment, making the transition to quantum statistical mechanics.

\section{Discussion and Conclusions}
In this note, we included the force that an inhomogeneous entropy might exert on the dynamics of particles in phase space. This is not an obvious step. It is a purely collective effect. So far, any such force has not yet been included in the dynamics of large numbers of particles and kinetic theory. 

Collective effects of this kind are known from ponderomotive forces and pressure forces. However, the inclusion of a separate ponderomotive force or a pressure force on the microscopic level is not necessary. Ponderomotive forces are taken care of by the interaction of the electromagnetic field with the particles. They arise from correlations. Similarly, the pressure force, which is directly proportional to the gradient of the particle density on any level, is already included in the evolution of the phase space density. 

The entropy force is different in the sense that it is not obvious that it is given by the gradient of density. The entropy force takes care of the chaotic disorder that is produced by the dynamics itself. Neither the pressure, nor the temperature account for it. Entropy is a function of the phase space density, not its moment like pressure and temperature. It is the cause of chaotic expansion of the phase space in the course of the dynamics. It therefore gives rise to a collective effect felt on the level of the microscopic phase space density. In many processes, this effect may be very small and negligible. This~will depend on scales. Elucidating those effects requires investigation of particular cases. The~present work presents the theory on which such attempts must be based. 

We derived the basic kinetic equation for the one particle phase space density including the global entropy force. Since this force is a global integral one, it has no microscopic source, while it affects the dynamics of the one-particle phase space density through the interaction Hamiltonian. This gives rise to the construction of a kinetic equation for the entropic phase space density, which turns out to be of similar structure, like the Liouville (Klimontovich) equation for the phase space density. There is, however, a fundamental difference in that the entropic kinetic equation is self-referential. It determines the dynamics of the entropy density in phase space by reference to the integrated entropy of the entire system itself. This is a very important finding because it shows that the entropy density in phase space evolves holographically. It is determined by itself. It takes care of its own evolution, which in addition depends on the evolution of the matter phase space density. This can be interpreted as a self-regulation of the evolution of entropy on the microscopic level, which takes care of the total produced entropy. According to this finding, any many-particle system that in the course of its dynamics generates entropy in an inhomogeneous and time-dependent way is subject to the entropy that controls its own evolution. This we feel is the most important insight we arrived at in our analysis. Entropy generates and controls itself in this highly nonlinear way already on the microscopic level of phase space density. We conjecture that these properties remain when including quantum effects in kinetic theory on the microscopic quantum level without the need to introduce a seed source of entropy.

In order to demonstrate an effect, we in the introductory part of this note treated the astrophysical example of a Schwarzschild black hole. This led us to the definition of the Schwarzschild constant, a universal constant, which is essentially the Planck force, which so far had been formally postulated without finding a physical interpretation. Its meaning lies in the Schwarzschild constant as the entropy force at the black hole horizon. Some of its implications we have discussed briefly. The entropy force at the horizon, being independent of charge, is of only one sign. It causes a repulsive force. This is similar to the gravitational force, though counteracting it. It thus in large space may compete with the gravitational attraction, where it may become kind of an anti-gravity. Investigation of its effect on the universal expansion might also be of interest, as also the role it may play in primordial black holes and planckions, which may have been generated in the early universe, the Big Bang, and if surviving, possibly due to the anti-gravity action of the entropy force, could provide all or part of the mysterious dark matter on the scales of clusters of galaxies. 
%%%%%%%%%%%%%%%%%%%%%%%%%%%%%%%%%%%%%%%%%%
\begin{acknowledgement}
This work was part of a brief Visiting Scientist Programme at the International Space Science Institute Bern. We acknowledge the interest of the ISSI directorate as well as the generous hospitality of the ISSI staff, in particular the assistance of the librarians Andrea Fischer and Irmela Schweitzer, and the Systems Administrator Saliba F. Saliba. 
\end{acknowledgement}

\end{document}